\documentclass{aa}  

\usepackage[]{graphicx}
\usepackage{amsmath}
\usepackage{amssymb}
\usepackage[separate-uncertainty=true,detect-weight]{siunitx}
\usepackage[]{hyperref}
\usepackage{cleveref}
\usepackage{caption}
\usepackage{subcaption}

\usepackage{microtype}
\usepackage{booktabs}
\usepackage{threeparttable}
\usepackage{enumerate}
\usepackage{multicol}
\usepackage{multirow}

\usepackage{comment}

\usepackage{longtable}
\usepackage{threeparttablex}
\usepackage{xcolor}
\usepackage{url}
\usepackage[normalem]{ulem}

\newcommand{\erosita}{eROSITA}
\newcommand{\srg}{\textit{SRG}}
\newcommand{\hst}{\textit{HST}}
\newcommand{\chandra}{\textit{Chandra}}
\newcommand{\rosat}{\textit{ROSAT}}
\newcommand{\erass}{\emph{eRASS:4}}
\newcommand{\newton}{\textit{XMM-Newton}}

\defcitealias{baldassare2022a}{B+22}

\DeclareSIUnit \pc {pc}
\DeclareSIUnit \parsec {parsec}
\DeclareSIUnit \arcsec {arcsec}
\DeclareSIUnit \pixel {pixel}
\DeclareSIUnit \pixels {pixels}
\DeclareSIUnit \Msun {M_{\odot}}
\DeclareSIUnit \smass {M_{\star}}
\DeclareSIUnit \dex {dex}
\DeclareSIUnit \mag {mag}
\DeclareSIUnit \yr {yr}
\DeclareSIUnit \angstrom {\mathrm{Å}}
\DeclareSIUnit \erg {erg}

\creflabelformat{equation}{#2#1#3}

\graphicspath{{./fig/}}

\begin{document} 

\title{Massive black holes in nuclear star clusters:}
\subtitle{Investigation with {\srg}/{\erosita} X-ray data}

\author{
  \href{https://www.orcid.org/0000-0001-8040-4088}{N.~Hoyer}\inst{1,2,3,4}\and
  \href{https://www.orcid.org/0000-0003-4054-7978}{R.~Arcodia\thanks{NASA Einstein fellow}}\inst{5,6}\and
  \href{https://www.orcid.org/0000-0002-6381-2052}{S.~Bonoli}\inst{1,7}\and
  \href{https://www.orcid.org/0000-0002-0761-0130}{A.~Merloni}\inst{5}\and
  \href{https://www.orcid.org/0000-0002-6922-2598}{N.~Neumayer}\inst{2}\and
  \href{https://www.orcid.org/0000-0001-8632-904X}{Y.~Zhang}\inst{5}\and
  \href{https://www.orcid.org/0000-0001-9200-1497}{J.~Comparat}\inst{5}
}
\institute{
  Donostia International Physics Center, Paseo Manuel de Lardizabal 4, E-20118 Donostia-San Sebasti{\'{a}}n, Spain\\\email{\href{mailto:nils.hoyer@dipc.org}{nils.hoyer@dipc.org}}\and
  Max-Planck-Institut f{\"{u}}r Astronomie, K{\"{o}}nigstuhl 17, D-69117 Heidelberg, Germany\and
  Institute of Astronomy, Pontificia Universidad Cat{\'{o}}lica de Chile, Avenida Vicu{\~{n}}a Mackena 4690, Santiago, Chile\and
  Universit{\"{a}}t Heidelberg, Seminarstrasse 2, D-69117 Heidelberg, Germany\and
  Max-Planck-Institut für extraterrestrische Physik (MPE), Giessenbachstrasse 1, D-85748 Garching, Germany\and
  MIT Kavli Institute for Astrophysics and Space Research, 70 Vassar Street, Cambridge, MA 02139, USA\and
  Ikerbasque, Basque Foundation for Science, E-48013 Bilbao, Spain
  }
\date{Received xx yy, 2023; accepted xx yy, 2023}

\abstract
{Massive black holes (MBHs) are typically hosted in the centres of massive galaxies but they appear to become rarer in lower mass galaxies, where  nuclear star clusters (NSCs) frequently appear instead. The transition region, where both an MBH and NSC can co-exist, has been poorly studied to date and only a few dozen galaxies are known to host them. One avenue for detecting new galaxies with both an MBH and NSC is to look for accretion signatures of MBHs.}
{Here, we use new {\srg/\erosita} all-sky survey {\erass} data to search for X-ray signatures of accreting MBHs in NSCs, while also investigating  their combined occupation fraction.}
{We collected more than \num{200} galaxies containing an NSC, spanning multiple orders in terms of galaxy stellar mass and morphological type, within the footprint of the German {\erosita} Consortium survey. We determined the expected X-ray contamination from binary stellar systems using the galaxy stellar mass and star formation rate as estimated from far-ultraviolet and mid-infrared emission.}
{We find significant detections for \num{18} galaxies (\SI{\sim 8.3}{\percent}), including one ultra-luminous X-ray source; however, only three galaxies (NGC{\,}2903, 4212, and 4639) have X-ray luminosities that are higher than the expected value from X-ray binaries, indicative of the presence of an MBH. In addition, the X-ray luminosity of six galaxies (NGC{\,}2903, 3384, 4321, 4365, 4639, and 4701) differs from previous studies and could indicate the presence of a variable active galactic nucleus. For NGC{\,}4701 specifically, we find a variation of X-ray flux within the {\erass} data set. Stacking X-ray non-detected galaxies in the dwarf regime ($M_{\star}^{\mathrm{gal}} \leq \SI{e9}{\Msun}$) results in luminosity upper limits of a few times \SI{e38}{\erg\per\second}.
The combined occupation fraction of accreting MBHs and NSCs becomes non-zero for galaxy masses above $\sim 10^{7.5}\, \si{\Msun}$ and this result is slightly elevated as compared to the literature data.}
{Our data extend, for the first time, towards the dwarf elliptical galaxy regime and identify promising MBH candidates for higher resolution follow-up observations. At most galaxy masses (and with the exception of three cases), the X-ray constraints are consistent with the expected emission from binary systems or an Eddington fraction of at most \SI{0.01}{\percent,} assuming a black holes mass of $10^{6.5} \, \si{\Msun}$. This work confirms the known complexities in similar-type of studies, while providing the appealing alternative of using X-ray survey data of in-depth observations of individual targets with higher resolution instruments.}


\maketitle

%
\section{Introduction}
\label{sec:introduction}

Since the first detections of massive compact objects in nearby galaxy centres almost forty years ago \citep{tonry1984b}, it has become evident that massive black holes (MBHs) occupy many nearby galaxy centres \citep[e.g.][]{kormendy1995a,magorrian1998a,tremaine2002a,kormendy2013a}.
This insight was made possible by significant advancements in the performance and capabilities of many ground-based facilities, including NIRC on Keck \citep{ghez1998a,filippenko2003a,walsh2012a}, SHARP on NTT \citep{genzel2000d,gillessen2009a}, GRAVITY \citep{gravity2017a,gravity2021a}, SAURON \citep{bacon2001c,vandenbosch2010a}, CFHT \citep{bender1996c,kormendy1997a}, SINFONI \citep{nowak2008a,rusli2011a,saglia2016a}, VLBI \citep{kuo2011a}, GEMINI/NIFS \citep{nguyen2018a,merrell2023b}, VLT \citep{marconi2001a}, and with the \textit{Hubble Space Telescope} \citep[{\hst}; e.g.][]{gebhardt2003a,devereux2003a,atkinson2005a,gültekin2009a,walsh2010a,nguyen2019a}, as well as improvements in dynamical models of galaxy centres \citep[e.g.][but see also \citealp{thater2022a} and references therein]{cappellari2004a,cappellari2020a,thater2019a,thater2022b}.
These measurements were only performed on massive galaxies as secure detections of MBHs towards the lowest galaxy masses become rare both because of weaker observational signatures and an apparent decline in the MBH occupation fraction, as suggested by observational \citep[e.g.][]{miller2015a,trump2015a,nguyen2018a} and theoretical \citep[e.g.][]{volonteri2003b,bellovary2011a,habouzit2017a,haidar2022a} studies.\footnote{See \citet{bustamente-rosell2021a} and \citet{regan2023a} for a discussion on a \SI{e6}{\Msun} MBH in the nearby Leo{\,}I ($M_{\star}^{\mathrm{gal}} \sim \SI{e7}{\Msun}$) dwarf galaxy.}
Despite numerous investigations \citep[see e.g.][for recent studies]{sharma2022a,beckmann2023a,spinoso2023a}, the functional shape and value of the decline of the occupation fraction from unity as a function of galaxy stellar (or halo) mass remain only loosely constrained.

Galaxy centres can also host dense stellar systems, known as nuclear star clusters (NSCs), which are more commonly found in the dwarf galaxy regime, occupying about \SI{80}{\percent} of $M_{\star}^{\mathrm{gal}} \sim \SI{e9}{\Msun}$ galaxies in the local universe \citep{sanchez-janssen2019a,neumayer2020a,hoyer2021a,ashok2023a}.
Contrary to MBHs, their occupation fraction rapidly declines in the most massive galaxies, where MBHs are most common, potentially due to interactions between the two objects \citep[e.g.][]{antonini2015b,arca-sedda2017b}  or tidal evaporation of progenitor clusters \citep{leaman2022a}.
Nevertheless, a transition region where both types of nuclei are present exists and includes, for example, the Milky Way \citep[e.g.][]{genzel2010b}.
As the functional shape of the MBH occupation fraction with respect to the host galaxy stellar mass is currently unclear, the extent of this transition region is unclear as well.

Due to observational constraints all firm MBH detections within NSCs are confined to relatively nearby galaxies (see e.g.\ Figure~\num{2} in \citealp{greene2020a} and the compilation of \citealp{neumayer2020a}) and are located in the NSC's centre with the exception of M{\,}31 \citep[e.g.][]{lauer1993a,bacon1994a,bacon2001a,bender2005a}.
Consequently, the total number of these systems is limited to a few dozen \citep[e.g.][]{neumayer2020a,nguyen2022a,thater2023a}, including ultra-compact dwarfs as previous NSCs of accreted galaxies \citep[e.g.][]{seth2014a,pfeffer2016a,ahn2017a,pechetti2022a}.
As we are now aware of more than 1000 nucleated galaxies \citep{munoz2015a,venhola2018a,sanchez-janssen2019a,carlsten2020a,habas2020a,poulain2021a,su2021a,hoyer2023a} and given the significant overlap between the NSC and MBH occupation fractions, we should expect a significantly higher number of galaxies with both an MBH and NSC.
While dynamical measurements are important to obtain reliable mass measurements for MBHs within NSCs, focusing on accretion signatures can help to identify larger samples out to higher distances, including dwarf galaxies where NSCs are most common \citep[e.g.][]{kauffmann2003c,baldassare2018a,birchall2020a,mezcua2020a,mezcua2023a,cann2023a}.

Accretion events onto MBHs from gas or stars via tidal disruption events \citep{rees1988c} leads to bright X-ray emission \citep[e.g.][]{komossa1999a,esquej2008a,maksym2010a} which can be used to study the mass of the black hole \citep[e.g.][]{mockler2019a} and potentially that of black hole binaries \citep{mockler2023a}.
Additionally, data from large-scale surveys was previously used to trace MBHs \citep[e.g.][]{miller2015b} and to constrain their occupation fraction \citep{miller2015a}.
One avenue to detect more MBHs in NSCs is to combine optical and X-ray data to detect and characterise the NSC and MBH, respectively, requiring an active galactic nucleus (AGN) that does not outshine the NSC in the optical regime.

Previous works have already taken advantage of combining various wavelength regimes \citep{seth2008a,baldassare2022a}, using, among other instruments, {\chandra} for X-rays.
Another approach compared to using high-resolution {\chandra} data is to perform a shallower wide-area survey, allowing us to study a greater number of NSCs in galaxies of various masses and morphologies.
The extended ROentgen Survey with an Imaging Telescope Array instrument \citep[{\erosita};][]{predehl2021a} aboard the \textit{Spectrum-Roentgen-Gamma} \citep[\srg;][]{sunyaev2021a} takes this approach and serves as an ideal laboratory for such a study.
The poorer resolution of {\erosita} operating in its survey mode (half-energy width of \SI{26}{\arcsecond}; \citealp{predehl2021a}) does not allow us to distinguish clearly between nuclear and off-nuclear emission as securely as {\chandra} but can still be used to detect MBH candidates for follow-up studies and to potentially probe MBH signatures in a large number of NSCs directly.
\begin{figure*}
    \centering
    \includegraphics[width=0.99\textwidth]{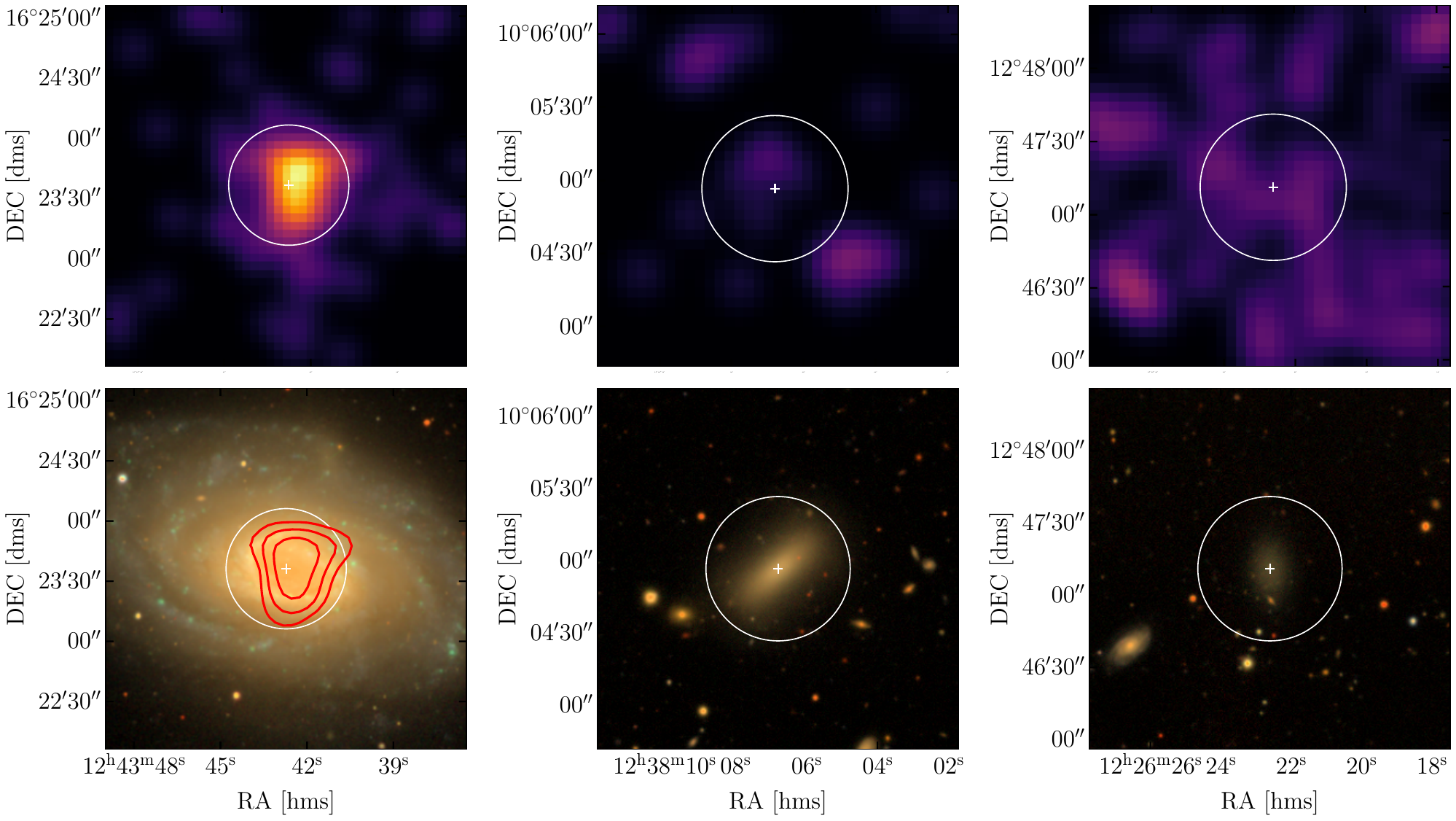}
    \caption{%
         Cutouts of {\erosita} {\erass} images (\emph{top panels}) and of the DESI Legacy Imaging Surveys (\emph{bottom panels}) Data Release 10 [Legacy Surveys / D.~Lang (Perimeter Institute)] of three galaxies in our sample: NGC 4651, IC 3602 and LEDA 40679.
         Both images are centered at the input optical coordinates of the NSC (white cross), with the \SI{30}{\arcsecond} aperture circle used for X-ray photometry highlighted in white.
         In case of an X-ray detection (\emph{left column}) X-ray contours are also overlayed to the optical image (red).
    }
    \label{fig:erass_example}
\end{figure*}

In this paper, we explore these possibilities using the cumulative data from {\erosita}'s already completed four all-sky surveys \citep[dubbed {\erass};][]{predehl2021a} to locate X-ray emission in a large sample of NSCs.
We introduce the {\erosita}, galaxy and literature data sets in \Cref{sec:data} and we analyse their properties in \Cref{sec:analysis}.
\Cref{sec:discussion} presents a discussion of the results and \Cref{sec:conclusions} gives the conclusions.

\section{Data}
\label{sec:data}

\subsection{Sample of nucleated galaxies}
\label{subsec:sample_of_nucleated_galaxies}

To generate an all-sky catalogue of nucleated galaxies, we first consider all galaxies up to a distance of \SI{100}{\mega\pc}, which are part of the HyperLEDA\footnote{\url{https://leda.univ-lyon1.fr/}} data base \citep{makarov2014a}, containing approximately \num{63000} objects.
Based on this catalogue, we search the \textit{Hubble Legacy Archive}\footnote{\url{https://hla.stsci.edu/}} for available high-resolution imaging data (\textit{Advanced Camera for Surveys}, \textit{Wide Field and Planetary Camera 2}, and \textit{Wide Field Camera 3}).
Based on these data, we assigned a nuclear classification to all galaxies, not taking into account previous classifications in the literature.
The HyperLEDA data base becomes incomplete towards the dwarf galaxy regime, which is why we add to the classified galaxy sample the data of \citet{denbrok2014a,munoz2015a,sanchez-janssen2019a,zanatta2021a,su2022a} for members of the Fornax, Virgo, and Coma galaxy clusters.
The combined catalogue contains \num{888} nucleated galaxies across the whole sky, which we used to cross-match with the German footprint of eROSITA.

\subsection{{\erosita} observations: {\erass}}
\label{subsec:erosita_observations}

We systematically extracted X-ray photometry at the input coordinates of the nucleated galaxies in the cumulative {\erass} images within the footprint of the German {\erosita} Consortium (i.e.\ Galactic longitudes between \num{179.944} and \num{359.944}).
This led to a starting sample of \num{239}/\num{888} galaxies, with mean exposure of $\sim \SI{418}{\second}$ and standard deviation of $\sim \SI{58}{\second}$ (see \Cref{fig:erass_example}) for three examples.
A detailed description of the methodology is presented in \citet{arcodia2023a} and we only outline here the basic steps.

X-ray counts were extracted between \num{0.2}-\SI{2.0}{\kilo\eV} within a circular aperture of \SI{30}{\arcsecond}, corresponding to $\sim \SI{75}{\percent}$ of the encircled energy fraction of {\erosita}'s point spread function in the adopted energy band.
The background contribution was estimated from an annulus with inner and outer radii of \SI{120}{\arcsecond} and \SI{360}{\arcsecond}, respectively.
Contaminating X-ray sources were masked following the prescription from \citet{comparat2023a} (Appendix~A).
For a small number of cases (\num{21}/\num{239}), more than 70 percent of the source aperture was masked out and the NSC was therefore excluded from the analysis.
Consequentially, the sample size of nucleated galaxies with extracted X-ray properties from the automated pipeline reduced down to $\num{239}-\num{21} = \num{218}$.

From {\erass} X-ray spectra were extracted from the same aperture using the \texttt{srctool} task in the {\erosita} Science Analysis Software System \citep[\texttt{eSASS},][]{brunner2022a}, with products version \texttt{020}.
The spectral analysis was performed with the Bayesian X-ray Analysis software (\texttt{BXA}) version \texttt{4.0.5.}\ \citep{buchner2014a}, which connects the nested sampling algorithm \texttt{UltraNest} \citep{buchner2019c,buchner2021a} with the fitting environment \texttt{XSPEC} version \texttt{12.12.0.}\ \citep{arnaud1996a}, in its Python version \texttt{PyXspec}.\footnote{The documentation for \texttt{PyXspec} can be found \href{https://heasarc.gsfc.nasa.gov/docs/xanadu/xspec/python/html/index.html}{here}.}
We adopted a simple power-law model (\texttt{zpowerlw}) with absorption fixed at the Galactic column density from HI4PI \citep{hi4pi2016a} and redshifted to rest-frame.
We quoted median, first, and 99th percentiles from the fit posteriors for fluxes and luminosity, unless otherwise stated.
In our approach, we followed the method from \citet{arcodia2023a} and adopted $P_{\mathrm{binom}} = \num{3e-4}$ as threshold for a significant detection, which corresponds to a spurious fraction of $\sim \SI{1}{\percent}$.
For non-detections ($P_{\mathrm{binom}} > \num{3e-4}$), we quoted upper limits using the 99th percentiles of the fit posteriors.

Potential individual sources of contamination from within the source aperture were treated a posteriori after visual inspection and were considered on a case-by-case basis.
For instance, we cross-matched our sample with the catalogue from \citet{walton2022a}, which compiled ultraluminous X-ray source (ULXs) candidates from \emph{XMM–Newton}, \emph{Swift-XRT}, and \emph{Chandra} data.
We manually masked out a handful of apertures with known ULX candidates and other obvious off-nuclear X-ray sources, whose centroid lies within the source aperture.
In some cases, this resulted in the NSC being non-detected after the masking:
the galaxy NGC{\,}4559 contains a known ULX \citep{walton2022a} and after its masking, the whole source aperture is masked-out and no products from the NSC can be analysed.
Therefore, after a visual inspection, the number of galaxies with extracted X-ray properties from the automated pipeline was \num{217}/\num{218}.
We provide in \Cref{tab:data} the properties of all \num{217} galaxies, derived as explained in the next subsections, with their measured eRASS:4 X-ray luminosities.

From this sample, we obtain that \num{18}/\num{217} targets are detected.
After computing  1$\sigma$ binomial uncertainties on this fraction from \citet{cameron2011a}, this results in a detection fraction of \num{8.3}$^{+2.3}_{-1.5} \, \si{\percent}$.

Finally, we performed a stacking analysis of non-detections following the methodology outlined in \citet{comparat2022a} and \citet{arcodia2023a}.
Around each galaxy, we retrieved a photon cube with the angular position, physical distance to the associated galaxy, exposure time, observed and emitted energy (shifted to the rest-frame of the galaxy), and the effective area for each photon.
Detected sources in the field were masked.
The cubes were merged for the desired sub-sample of non-detected galaxies.
We took a weighted mean of all events within $\SI{10}{\kilo\pc}$ using the weight described in Equ.~3 of \citet{comparat2022a} to obtain the surface brightness.
We estimated the background surface brightness by repeating the procedure with events located at a distance between \num{15} and \SI{50}{\kilo\pc}.
Finally, we subtracted the background from the mean surface brightness (within a \SI{10}{\kilo\pc} radius) and converted it to a luminosity.
\begin{table*}
  \tiny
  \caption{%
    Literature and {\erosita} {\erass} data for our sample of nucleated galaxies.
    The complete data table is available for download online.
  }
  \begin{center}
  \scalebox{0.7}{
    \begin{threeparttable}
      \begin{tabular}{%
        l
        l
        l
        l
        l
        l
        l
        l
        l
        l
        l
        l
        l
      }
        \toprule
        \multicolumn{1}{c}{Galaxy} & \multicolumn{1}{c}{RA} & \multicolumn{1}{c}{DEC} & \multicolumn{1}{c}{$m-M$} & \multicolumn{1}{c}{$T$} & \multicolumn{1}{c}{$\log_{10} \, M_{\star}^{\mathrm{gal}}$} & \multicolumn{1}{c}{$\log_{10} \mathrm{SFR}$ \tnote{(a)}} & \multicolumn{1}{c}{$L_{X, \, \mathrm{median}}^{\mathrm{2-10} \, \si{\kilo\eV}}$} & \multicolumn{1}{c}{$L_{X, \, \mathrm{1}^{\mathrm{st}}}^{\mathrm{2-10} \, \si{\kilo\eV}}$} & \multicolumn{1}{c}{$L_{X, \, \mathrm{99}^{\mathrm{th}}}^{\mathrm{2-10} \, \si{\kilo\eV}}$} & \multicolumn{1}{c}{$L_{X}^{\mathrm{cont}}$ \tnote{(b)}} & \multicolumn{1}{c}{$t_{\mathrm{exp}}$} & \multicolumn{1}{c}{$P_{\mathrm{binom}}$} \\
        \cmidrule(lr){2-2}
        \cmidrule(lr){3-3}
        \cmidrule(lr){4-4}
        \cmidrule(lr){6-6}
        \cmidrule(lr){7-7}
        \cmidrule(lr){8-8}
        \cmidrule(lr){9-9}
        \cmidrule(lr){10-10}
        \cmidrule(lr){11-11}
        \cmidrule(lr){12-12}
        & \multicolumn{1}{c}{[deg]} & \multicolumn{1}{c}{[deg]} & \multicolumn{1}{c}{[mag]} & & \multicolumn{1}{c}{[\si{\Msun}]} & \multicolumn{1}{c}{[\si{\Msun\per\yr}]} & \multicolumn{1}{c}{[\si{\erg\per\second}]} & \multicolumn{1}{c}{[\si{\erg\per\second}]} & \multicolumn{1}{c}{[\si{\erg\per\second}]} & \multicolumn{1}{c}{[\si{\erg\per\second}]} & \multicolumn{1}{c}{[\si{\second}]} & \\
        \multicolumn{1}{c}{(1)} & \multicolumn{1}{c}{(2)} & \multicolumn{1}{c}{(3)} & \multicolumn{1}{c}{(4)} & \multicolumn{1}{c}{(5)} & \multicolumn{1}{c}{(6)} & \multicolumn{1}{c}{(7)} & \multicolumn{1}{c}{(8)} & \multicolumn{1}{c}{(9)} & \multicolumn{1}{c}{(10)} & \multicolumn{1}{c}{(11)} & \multicolumn{1}{c}{(12)} & \multicolumn{1}{c}{(13)} \\
        \midrule
        {BTS{\,}76} & {$179.68375$} & {$27.58500$} & {$30.50$} & {$10.0 \pm 2.0$} & {$7.53 \pm 0.16$} & {$-3.47$} & {-{-}} & {-{-}} & {$3.97 \times 10^{38}$} & {$(6.2 \pm 2.3) \times 10^{36}$} & {$326$} & {0.06078} \\
        {DDO{\,}084} & {$160.67458$} & {$34.44889$} & {$29.99$} & {$9.8 \pm 0.6$} & {$7.65 \pm 0.35$} & {$-3.84$} & {-{-}} & {-{-}} & {$1.07 \times 10^{38}$} & {$(8.1 \pm 6.5) \times 10^{36}$} & {$325$} & {0.39909} \\
        {DDO{\,}088} & {$161.84292$} & {$14.07028$} & {$29.44$} & {$8.9 \pm 0.3$} & {$7.88 \pm 0.15$} & {N/A} & {-{-}} & {-{-}} & {$6.54 \times 10^{38}$} & {$(1.37 \pm 0.48) \times 10^{37}$} & {$282$} & {$0.04318$} \\
        {dw{\,}1048+13} & {$162.14917$} & {$13.06000$} & {$30.13 \pm 0.12$} & {$-2.0 \pm 1.0$} & {$6.71 \pm 0.43$} & {N/A} & {-{-}} & {-{-}} & {$4.25 \times 10^{38}$} & {$(9.3 \pm 9.2) \times 10^{35}$} & {$297$} & {$0.40543$} \\
        {dw{\,}1049+12b} & {$162.35833$} & {$12.55250$} & {$30.17 \pm 0.06$} & {$-2.0 \pm 1.0$} & {$7.55 \pm 0.06$} & {N/A} & {-{-}} & {-{-}} & {$1.84 \times 10^{38}$} & {$(6.37 \pm 0.81) \times 10^{36}$} & {$275$} & {$0.76436$} \\
        {\ldots} & {\ldots} & {\ldots} & {\ldots} & {\ldots} & {\ldots} & {\ldots} & {\ldots} & {\ldots} & {\ldots} & {\ldots} & {\ldots} & {\ldots} \\
        \bottomrule
      \end{tabular}
      \begin{tablenotes}
        \small
        \item[] \hspace{-5pt}Columns: (1) galaxy name; (2) Right ascension; (3) Declination; (4) Distance modulus; (5) Hubble type; (6) Galaxy stellar mass; (7) Star formation rate; (8) Median X-ray luminosity in the \num{2}-\SI{10}{\kilo\eV} band; (9) \& (10) 1$^{\mathrm{st}}$ and 99$^{\mathrm{th}}$ percentiles of the X-ray luminosity; (11) Expected X-ray luminosity due to binary systems; (12) Exposure time; (13) Binomial probability that the detection is a background fluctuation.
        \item[(a)] The star formation rate is set to zero for dwarf elliptical galaxies in the core of the Virgo cluster.
        \item[(b)] If no star formation rate could be determined, the estimated X-ray luminosity due to binaries is a lower limit.
      \end{tablenotes}
    \end{threeparttable}
  }
  \end{center}
  \label{tab:data}
\end{table*}

\subsection{NSC and galaxy parameters}
\label{subsec:nsc_galaxy_parameters}

The imaging data from {\hst} used to classify galaxies are inhomogeneous with different spatial resolutions and available filters.
Instead of deriving new NSC parameter estimates from these data, we looked for available values from the literature.
As a consequence, not all nucleated galaxies of our sample have available NSC properties (seen in \Cref{fig:nscmass_galmass,fig:nscreff_nscmass} below).
More specifically, we searched for available NSC parameters for nucleated galaxies within the
Local Volume \citep[$d \lesssim \SI{11}{\mega\pc}$;][]{seth2006a,georgiev2009b,graham2009b,baldassare2014a,schoedel2014a,calzetti2015a,carson2015a,crnojevic2016a,nguyen2017a,baumgardt2018b,nguyen2018a,bellazzini2020a,pechetti2020a,hoyer2023a}, the field environment outside the Local Volume \citep{georgiev2014a,georgiev2016a}, and the Virgo galaxy cluster \citep{sanchez-janssen2019a}.
Nucleated galaxies in the Coma and Fornax galaxy clusters are not part of our data set.

Galaxy stellar masses were determined as presented in the next subsection.
Morphological type values are adopted from HyperLEDA and for the data sample of \citet{sanchez-janssen2019a}, we assumed a value of \num{-5}, corresponding to elliptical galaxies.

\subsubsection{Galaxy stellar masses}
\label{subsubsec:galaxy_stellar_masses}

To compute galaxy stellar masses, we used three different tracers:
(1) the $B-V$ colour, (2) the $g-r$ colour, and (3) the $K$-band luminosity.
First, we obtained the photometric parameters and distance estimates from both the HyperLEDA and NED data bases.
We then computed the aforementioned colours and $B$-, $g$-, and $K$-band luminosities using the absolute magnitude of the Sun\footnote{URL: \url{http://mips.as.arizona.edu/~cnaw/sun.html}} \citep{willmer2018a}, accounting for Galactic extinction via the re-calibrated version of the \citet{schlegel1998a} extinction maps \citep{schlafly2011a}, assuming $R_{V} = 3.1$ \citep{fitzpatrick1999a}.
Internal extinction was not taken into account.
The mass-to-light ratios for the different colours and $K$-band luminosity were taken from \citet{mcgaugh2014a} and \citet{du2020a}, which give re-calibrated versions of the original relations by \citet{bell2003a,portinari2004a,zibetti2009a,into2013a,roediger2015a}.
The uncertainties on the final stellar mass estimates were based on those of the photometric parameters, the distance estimate, the absolute magnitude of the Sun (assumed to be \SI{0.04}{\mag}), and the stellar mass-to-light relation (assumed to be \SI{0.3}{\dex}).
Usually, the latter one dominates over all other uncertainties.

For the dwarf galaxies in the Virgo cluster, we directly took the mass estimates from \citet{sanchez-janssen2019a}, which are based on fits to the spectral energy distributions and overall match to the other three approaches outlined above \citep{hoyer2021a}.
Their stellar mass estimates lack an uncertainty which is why we assumed a value of \SI{0.3}{\dex}.

\subsection{Literature X-ray data}
\label{subsec:literature_x-ray_data}

As pointed out in the Introduction, previous work investigated the X-ray emissivity of NSCs in search for MBHs.
Based on optical spectroscopy as well as radio and X-ray data (via {\chandra}, {\rosat}, and {\newton}), \citet{seth2008a} found X-ray emission indicative of the presence of MBHs consistent with the position of NSCs in a sample of \num{176} early- and late-type galaxies.
Most recently, \citet{baldassare2022a} used data from the {\chandra} X-ray observatory to search for such signatures in \num{108} nearby ($d \lesssim \SI{40}{\mega\pc}$) nuclei from the galaxy sample of \citet{georgiev2014a}, which is composed of massive late-type galaxies.
They classified \num{29} targets as having significant X-ray emission and, thus, harbouring AGN.

Some other studies investigated the X-ray luminosity of the central region of galaxies without taking into account their nuclear classification.
Here, we took into account data from \citet{she2017a} and \citet{ohlson2023a}, which used archival {\chandra} data.

We extracted fluxes and luminosities in the \num{0.5}-\SI{7}{\kilo\eV} range to compare with \citet{ohlson2023a} and in the \num{2}-\SI{10}{\kilo\eV} range to compare with \citet{baldassare2022a}.
We note that compared to {\chandra}, {\erosita} is most sensitive in the \num{0.2}-\SI{2.3}{\kilo\eV} energy band \citep{predehl2021a}.
From \citet{ohlson2023a} we used their luminosity values for a circular aperture with a radius of \SI{3}{\arcsecond} to better match the PSF of {\erosita}.
\citet{baldassare2022a} gives luminosities in both bands and we confirmed that the results (given in \Cref{subsec:x-ray_variable_sources}) remain unchanged when changing to the \num{0.5}-\SI{7}{\kilo\eV} band.

\section{Analysis}
\label{sec:analysis}

\subsection{X-ray contamination from binaries}
\label{subsec:x-ray_contamination_from_binaries}

Both low- and high-mass X-ray binaries, namely,\ binary stellar systems composed of a donor and either a neutron star or stellar mass black hole, can significantly contribute to a galaxy's total X-ray luminosity \citep[e.g.][]{iwasawa2009a}, sometimes rivalling that of the AGN \citep[e.g.][]{lehmer2010b}.
This contribution is especially important for our analysis given the size of {\erosita}'s PSF (half-energy width of approximately \SI{30}{\arcsec}; \citealp{predehl2021a}).
The formation of low-mass X-ray binaries (LMXBs) typically takes \num{1}-\SI{10}{\giga\yr} \citep{verbunt1995a} as we require stellar evolution to first produce a neutron star, which then has to dynamically enter into a binary system with a donor.
The collective X-ray luminosity of these systems in older disks and bulges is related to the stellar mass of the galaxy \citep[][]{gilfanov2004b} via $L_{X}^{\textrm{LMXB}} = \alpha \times M_{\star}^{\mathrm{gal}}$ \citep[e.g.][]{colbert2004a,lehmer2010b}, where $\log_{10} \, \alpha = 29.15^{+0.07}_{-0.06} \; \si{\erg\per\second\per\Msun}$ \citep{lehmer2019a}.

In contrast, high-mass X-ray binaries (HMXBs) require a stellar mass black hole and their X-ray emission is related to the stellar evolution timescale of the massive donor star, resulting in a luminous phase about \SI{100}{\mega\yr} after formation of the binary \citep{verbunt1995a}.
Due to the high-mass of the donor and its short lifetime, the X-ray luminosity is related to the star formation rate ($\mathrm{SFR}$) of the host galaxy via $L_{X}^{\textrm{HMXB}} = \beta \times \mathrm{SFR}$ \citep[e.g.][]{grimm2003b}, where $\log_{10} \beta = 39.73^{+0.15}_{-0.10} \; \si{\erg\per\second} \, (\si{\Msun\per\yr})^{-1}$ \citep{lehmer2019a}.

To estimate the current star formation rate of our sample, which we assume to be constant over the last \SI{100}{\mega\yr} (i.e.\ no star bursts or quenching effects from tidal interactions or bright AGN), we used the correlation from \citet[][but see also from \citealp{kennicutt2012a}, then]{hao2011a} relating the emission in the far-ultraviolet ($L_{\mathrm{FUV}}$) with the mid-infrared ($L_{\mathrm{MIR}}$) and star formation rate via:
\begin{equation}
    \log_{10} \, \mathrm{SFR} = \log_{10} \big(L_{\mathrm{FUV}} + 3.89 \times L_{\mathrm{MIR}} \big) - \num{43.35} \, .
\end{equation}
We took this approach, contrary to, for instance, an estimation via X-rays \citep{colbert2004a,symeonidis2011a,riccio2023b}, due to the availability and homogeneity of the available luminosities:
to estimate the far-ultraviolet luminosity, we used the publicly available data from \textit{Galex} \citep{morrissey2007a}.
For the mid-infrared luminosity, we used the \textit{AllWISE} \textit{W4} \citep{wright2010a,cutri2013a} or \textit{Spitzer} \textit{MIPS} \citep{rieke2004a} magnitudes.
For dwarf elliptical galaxies in the Virgo cluster, we assumed that no star formation occured over the last few hundred \si{\mega\yr} and that the expected X-ray binary contamination is solely produced by LMXBs.

After computing the galaxy stellar mass and star formation rates, we determined the luminosities of both classes of binary systems.
The expected contamination by X-ray binary systems is the sum of the two components, $L_{X}^{\mathrm{bin}} = L_{X}^{\mathrm{LMXB}} + L_{X}^{\mathrm{HMXB}}$.
Objects that were detected above this expected X-ray binary emission could indicate the presence of an MBH (however, see \Cref{subsec:caveats} for caveats).

\subsection{Galaxy properties of X-ray detected sources}
\label{subsec:properties_of_x-ray_detected_sources}

\begin{figure}
    \centering
    \includegraphics[width=\columnwidth]{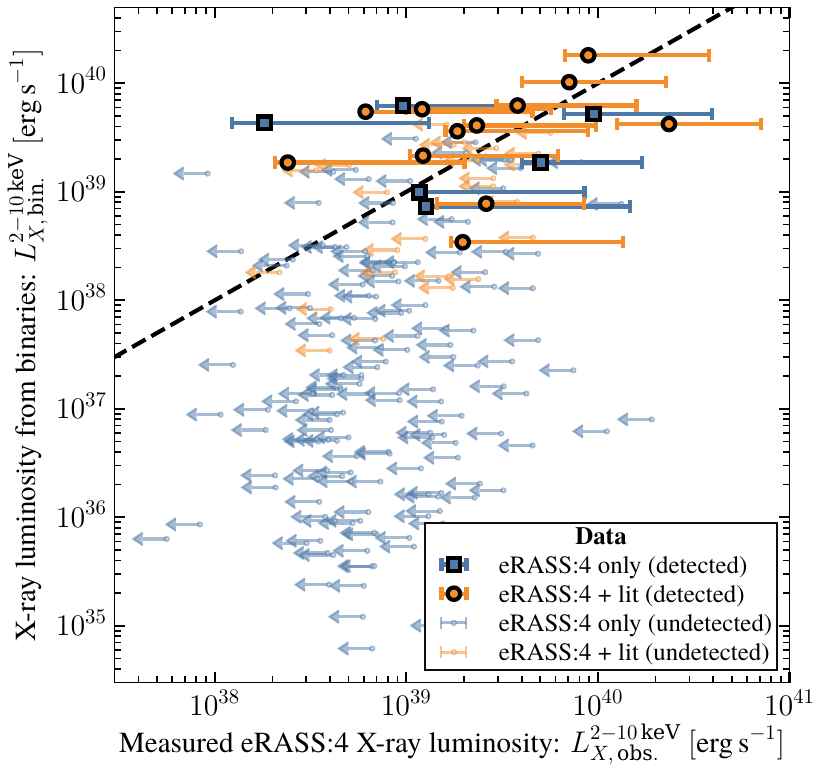}
    \caption{%
        Expected X-ray luminosity from the combined high- and low-mass binary population in the \num{2}-\SI{10}{\kilo\eV} range \big($L_{X, \, \mathrm{bin}}^{2\mathrm{-}10 \, \mathrm{keV}}$\big) versus the measurements $\big( L_{X, \, \mathrm{obs}}^{2\mathrm{-}10 \, \mathrm{keV}} \big)$.
        We show the detected galaxies ($P_{\mathrm{binom}} \leq \num{3e-4}$) in the {\erass} footprint with blue squares.
        For non-detections, we plot upper limits using the 99th percentile of the X-ray luminosity (blue arrows).
        Additionally, we highlight the {\erass} luminosities of galaxies with available literature X-ray data (see \Cref{subsec:literature_x-ray_data}; orange circles).
        These include data from \citet{seth2008a,she2017a,baldassare2022a,ohlson2023a}.
        Some of the galaxies are not detected in the {\erass} data but have secure measurements from other instruments available in the literature (orange arrows).
        A more detailed comparison between the {\erass} and literature data is presented in \Cref{subsec:x-ray_variable_sources}.
    }
    \label{fig:lumexp_lumobs}
\end{figure}

We compared the expected X-ray luminosity from the binary populations in the \num{2}-\SI{10}{\kilo\eV} range with the {\erass} data in \Cref{fig:lumexp_lumobs}, distinguishing between X-ray detected sources ($P_{\mathrm{binom}} \leq \num{3e-4}$) and undetected sources ($P_{\mathrm{binom}} > \num{3e-4}$, shown as upper limits).
We also distinguished between objects only detected with {\erosita} and the ones also detected with other instruments taken from \citet{seth2008a,she2017a,baldassare2022a} or \citet{ohlson2023a}.
All significant detections in the {\erass} data have measured X-ray luminosities of $L_{X, \, \mathrm{obs}}^{\mathrm{2-10} \, \si{\kilo\eV}} > \SI{e38}{\erg\per\second}$ and similarly high expected luminosities from the galaxies LMXBs and HMXBs.
Only three galaxies in our sample (NGC{\,}2903, NGC{\,}4212, and NGC{\,}4639) have measured luminosities greater than the expected values from binaries at $3\sigma$ confidence.
Some of the X-ray detected sources are also measured below the expected value which may be related to uncertainties in the estimates of the galaxy-only predictions.
We will discuss this observation further in \Cref{subsec:caveats}.
Below $L_{X, \, \mathrm{obs}}^{\mathrm{2-10} \, \si{\kilo\eV}} \sim \SI{5e38}{\erg\per\second}$ we find no significant emission and can only determine upper limits.
Some of the NSCs with X-ray upper limits in \Cref{fig:lumexp_lumobs} reside in dwarf elliptical galaxies in the core of the Virgo cluster (\num{29}/\num{117} galaxies at $M_{\star}^{\mathrm{gal}} \leq \SI{e9}{\Msun}$).
A lack of photometric data in the literature makes it challenging to determine star formation rates and we assumed that no star formation takes place for these objects.
While this assumption may be justified for the dwarf galaxy sample, the presented  X-ray luminosity from binaries remains only a lower limit.

\begin{figure}
    \centering
    \includegraphics[width=\columnwidth]{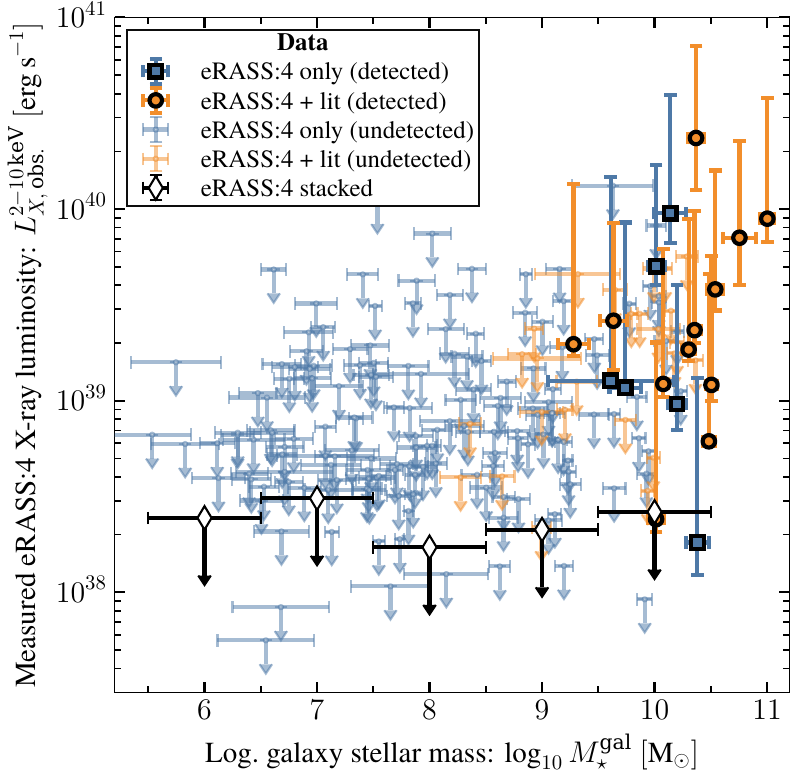}
    \caption{%
        Measured X-ray luminosity in the \num{2}-\SI{10}{\kilo\eV} range ($L_{X, \, \mathrm{obs}}^{\mathrm{2-10} \, \si{\kilo\eV}}$) versus galaxy stellar mass ($\log_{10} \, M_{\star}^{\mathrm{gal}}$).
        The markers, colour-coding and literature references are all the same as in \Cref{fig:lumexp_lumobs}.
        We stack non-detections (upper limits, blue and orange arrows) in mass bins of \SI{1}{\dex} width starting at a galaxy mass of $10^{5.5} \, \si{\Msun}$ to set tighter constraints on the X-ray emission in dwarf galaxies.
        Only galaxies above galaxy masses of approximately $10^{9.3} \, \si{\Msun}$ are detected in the {\erass} data.
    }
    \label{fig:lumobs_galmass}
\end{figure}

\Cref{fig:lumobs_galmass} shows the distribution of measured X-ray luminosity versus galaxy stellar mass.
To further constrain the emission in the dwarf galaxies, we stack non-detected galaxies within bins of stellar mass of \SI{1}{\dex} starting at $M_{\star}^{\mathrm{gal}} = 10^{5.5} \, \si{\Msun}$ until $10^{10.5} \, \si{\Msun}$ (see Section~\ref{subsec:erosita_observations}).
None of the stacked X-ray images contains signals above background level, with upper limits of the order of $\SI{2e38}{\erg\per\second}$ found in each stellar mass bin.
Estimating the expected X-ray luminosity from LMXBs with these galaxy stellar masses reveals that these upper limits are either matching or higher than the expected values; so, while we cannot rule out the presence of MBHs with low Eddington fractions (see \Cref{subsec:presence_of_massive_black_holes} for more discussion below), these galaxies do not contain X-ray bright AGN.

From the low-mass towards the high-mass end in \Cref{fig:lumobs_galmass} the X-ray luminosity of the significant detections appears to increase, starting around $M_{\star}^{\mathrm{gal}} \sim \SI{e10}{\Msun}$.
While this increase could be related to the increasing contribution from the nuclear emission, as seen for a few sources in \Cref{fig:lumexp_lumobs}, most of the detected sources feature the expected luminosities from X-ray binary systems.
Therefore, it appears plausible that this increase is mostly related to the increasing strength of LMXBs and not due to the presence of AGN.

\subsection{NSC properties of X-ray detected sources}
\label{subsubsec:nsc_properties}

\begin{figure}
    \centering
    \includegraphics[width=\columnwidth]{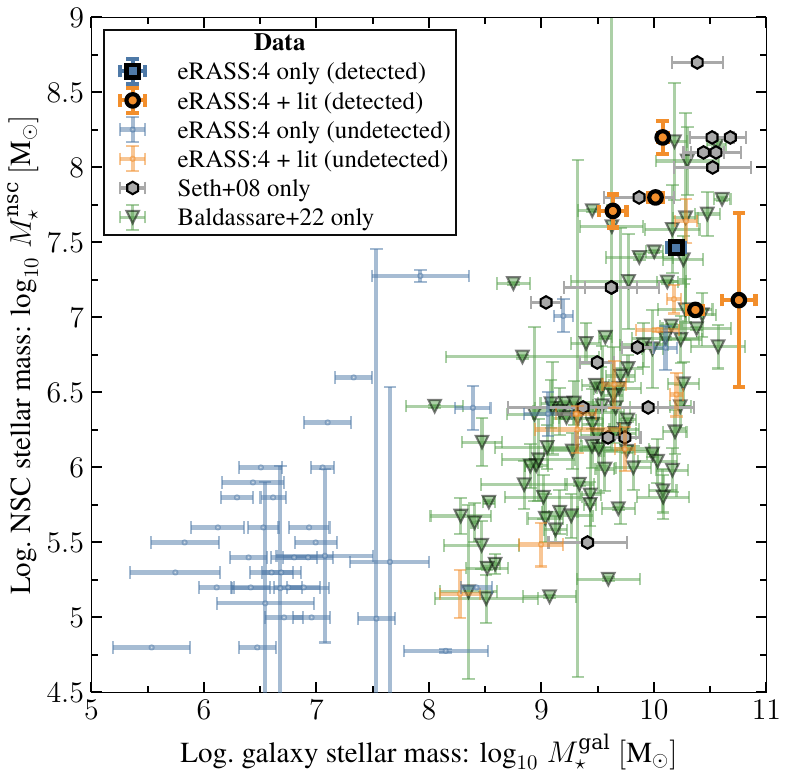}
    \caption{%
        Nuclear star cluster ($\log_{10} \, M_{\star}^{\mathrm{nsc}}$) versus host galaxy stellar mass ($\log_{10} \, M_{\star}^{\mathrm{gal}}$).
        We show X-ray detected NSCs with full colour, distinguishing between sources only detected with {\erosita} (blue) and sources also detected with other instruments (orange).
        A fainter shade is used for non-detection in the {\erass} data.
        In addition, we show NSCs analysed by \citet[][grey hexagons]{seth2008a} and \citet{baldassare2022a} (green triangles) for NSCs with X-ray emission outside the {\erass} footprint (or with significant contamination).
        A lack of literature data for NSC properties limits the included data set.
    }
    \label{fig:nscmass_galmass}
\end{figure}

Regarding NSC properties, we show the NSC versus host galaxy stellar mass relation \citep[e.g.][]{georgiev2016a,neumayer2020a,ashok2023a} in \Cref{fig:nscmass_galmass}.
We complement our data with the sample of \citet{baldassare2022a}, which is based on the data of \citet{georgiev2014a}.
Similar to the previous observation for galaxy masses in \Cref{fig:lumobs_galmass}, only the most massive NSCs are X-ray detected, above $M_{\star}^{\mathrm{nsc}} \sim \SI{e7}{\Msun}$.
Other lower-mass NSCs are not detected, including objects, which were detected previously with {\chandra} down to $M_{\star}^{\mathrm{nsc}} \gtrsim \SI{e5}{\Msun}$.

\begin{figure}
    \centering
    \includegraphics[width=\columnwidth]{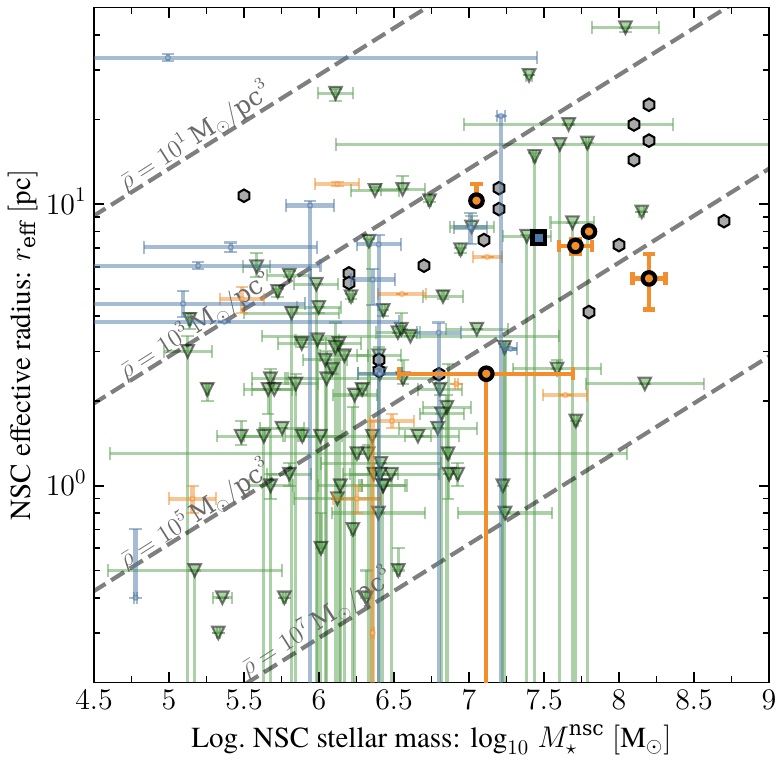}
    \caption{%
        Nuclear star cluster effective radius ($r_{\mathrm{eff}}$) versus stellar mass ($\log_{10} \, M_{\star}^{\mathrm{nsc}}$).
        The colour-coding and symbols are the same as in \Cref{fig:nscmass_galmass}.
        Note that compared to \Cref{fig:nscmass_galmass} fewer NSCs are shown because of a lack of measured effective radii for the dwarf galaxy sample in the core of the Virgo cluster \citep{sanchez-janssen2019a,ferrarese2020a}.
        Uncertainties are omitted for clarity.
    }
    \label{fig:nscreff_nscmass}
\end{figure}

\Cref{fig:nscreff_nscmass} shows the NSC effective radius versus stellar mass plane.
The {\erass} detected NSCs have both high stellar mass and large radii, as is expected from scaling relations \citep{georgiev2016a,neumayer2020a,ashok2023a}.
The average half-mass density of these objects, $\overline{\rho}$, is consistent with other X-ray detected but lower-mass NSCs in the literature, falling between $\SI{e4}{\Msun\per\pc\cubed} \lesssim \bar{\rho} \lesssim \SI{e6}{\Msun\per\pc\cubed}$.

\citet{baldassare2022a} also investigated the properties of their NSC sample distinguishing between X-ray luminosities likely originating from a massive black hole and X-ray binaries.
Due to our limited sample of new detections, we refer to their study for further discussion with NSC properties.

\subsection{X-ray variable sources}
\label{subsec:x-ray_variable_sources}

\begin{figure}
    \centering
    \includegraphics[width=\columnwidth]{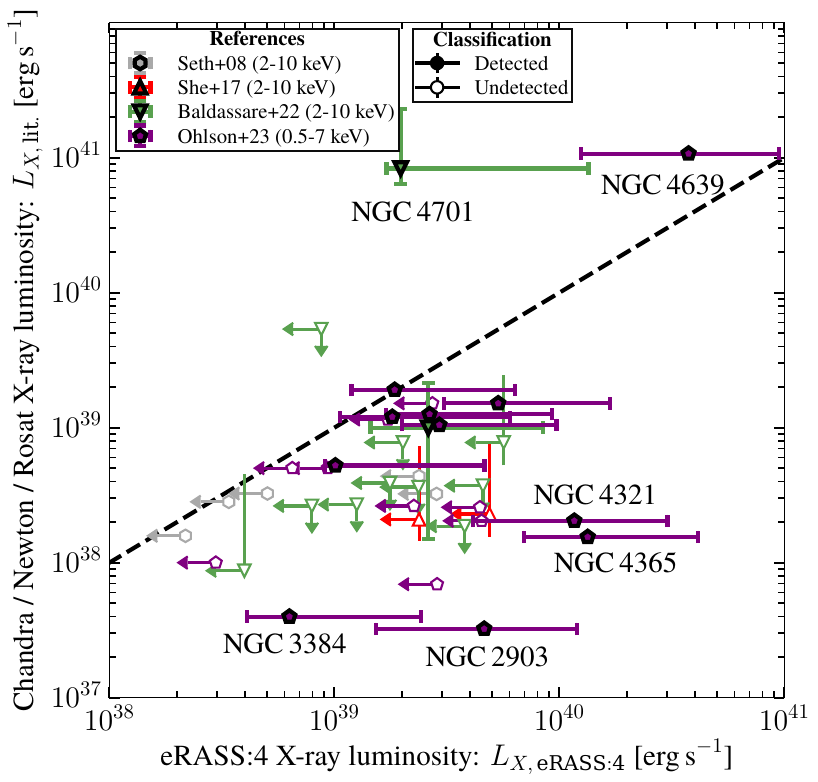}
    \caption{%
    Comparison of literature {\chandra}, {\newton}, or {\rosat} X-ray luminosity versus {\erass} values.
    Full colour-coded symbols show X-ray detections in {\erass}.
    The dashed line gives the one-to-one values.
    We compare to \citet{seth2008a}, \citet{she2017a} and \citet{baldassare2022a} using the \num{2} to \SI{10}{\kilo\eV} band.
    \citet{ohlson2023a} give {\chandra} X-ray data in the \num{0.5} to \SI{7}{\kilo\eV} band.
    Galaxies, which likely show some X-ray variability, are specifically named.
    }
    \label{fig:lumcomp}
\end{figure}

We compare in \Cref{fig:lumcomp} the X-ray luminosity of nucleated galaxies in the German footprint of the {\erass} data with matches in {\chandra}, {\rosat}, or {\newton} data from \citet{seth2008a}, \citet{she2017a}, \citet{baldassare2022a}, and \citet{ohlson2023a}, as introduced in \Cref{subsec:literature_x-ray_data}.
Given their upper limits, most of the {\erass} values are in agreement with the literature.
However, six other galaxies (NGC{\,}2903, 3384, 4321, 4365, 4639, and 4701) have values not in agreement with the literature.

For one galaxy, NGC{\,}2903, there exist literature data from both \citet{baldassare2022a} and \citet{ohlson2023a}.
Although the same data were analysed, they quote fluxes differing by a factor $\sim 100$, which is likely due to the difference between catalogue fluxes \citep{evans2010a,ohlson2023a} and those estimated through aperture photometry \citep{baldassare2022a}.
We computed the X-ray fluxes through spectral analysis and obtained that the {\erass} spectrum is sufficiently well described by a simple power-law with photon index $2.08 \pm 0.20$, although a more complex spectral model would be most likely required with higher count statistics in the \num{2}-\SI{10}{\kilo\eV} band.
Based on this, we are not able to infer whether the difference between {\erosita} and {\chandra} values is due to intrinsic variability or differences in the flux estimate methods.

We investigate this object further by looking at the X-ray luminosity in each {\erosita} survey to find that it was detected in \textit{eRASS2} (with $L_{X, \, \mathrm{obs}}^{\mathrm{0.2}-\mathrm{2.0} \, \si{\kilo\eV}} \sim \SI{1.1e40}{\erg\per\second}$) but not in any other individual image.
This indicates that NGC{\,}2903 likely hosts an AGN, which is variable on time scales shorter than six months (i.e.\ the time between all-sky scans by {\erosita}).

For the other five galaxies, we find no significant signs of X-ray variability within the {\erass} data.
This could indicate that the inconsistency detected here, if not caused by any differences in analysis strategy between our approach and the one of \citet{ohlson2023a}, occurs on time scales longer than six months but shorter than the time difference between the {\chandra} and {\erosita} observations of a few years.

\section{Discussion}
\label{sec:discussion}

\subsection{Evaluating the presence of massive black holes}
\label{subsec:presence_of_massive_black_holes}

\begin{figure}
    \centering
    \includegraphics[width=\columnwidth]{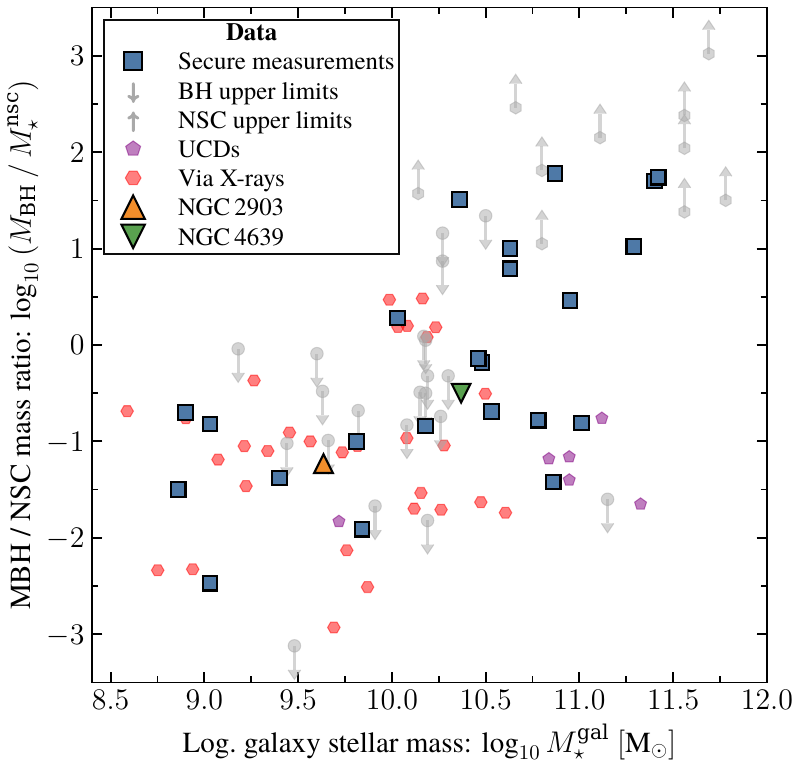}
    \caption{%
        Mass ratio of the SMBH and NSC stellar mass ($\log_{10} \, M_{\mathrm{BH}} / M_{\star}^{\mathrm{nsc}}$) versus the host galaxy stellar mass.
        We show the compiled data of \citet{neumayer2020a} and add to that data for NGC{\,}1336 \citep{saulder2016a,fahrion2019b,thater2023a}, NGC{\,}3593 \citep{bertola1996a,nguyen2022a}, and four galaxies from \citet{ashok2023a}, all shown with blue squares.
        The sample of ultra-compact dwarf galaxies (UCDs, purple pentagons) include the compilation of \citet{neumayer2020a} and the additional B023-G078, M{\,}31's most massive globular cluster, from \citet{pechetti2022a}.
        We estimate the most likely previous UCD's host galaxy mass using the $M_{\star}^{\mathrm{nsc}}$-$M_{\star}^{\mathrm{gal}}$ relation of \citet{neumayer2020a} for massive galaxies (their Equation \num{2}).
        Additionally, we add nucleated galaxies with significant X-ray excess, hinting at the existence of a massive black hole, by using a black hole mass estimate from \citet{she2017a} and NSC mass estimates from \citet{georgiev2016a}, shown with red hexagons.
        We add to this last group the data points of NGC{\,}2903 and NGC{\,}4639, which could host massive black holes based on their X-ray variability (see \Cref{fig:lumcomp}) with an orange and green triangle, respectively.
    }
    \label{fig:mass_ratio_galmass}
\end{figure}

Most of the X-ray detected NSCs in {\erass} are in agreement with the expected luminosity from binary system, therefore, we are not able to unambiguously associate it with an AGN.
Three NSCs (NGC{\,}2903, NGC{\,}4212, and NGC{\,}4639) have emission above the expected value and this supports the presence of an AGN.
For NGC{\,}4212 the NSC properties are not known and no secure black hole measurement exists.
For the other two galaxies, \citet{she2017a} estimates the black hole mass using the $M_{\mathrm{BH}}-\sigma$ relation from \citet{kormendy2013a} to find $\log_{10} \, M_{\mathrm{BH}}^{2903} \,/\, \mathrm{M}_{\odot} = 6.48^{+0.10}_{-0.10}$ and $\log_{10} \, M_{\mathrm{BH}}^{4639} \,/\, \mathrm{M}_{\odot} = 6.65^{+0.09}_{-0.09}$, resulting in $\lambda_{\mathrm{Edd}}^{2903} \sim \num{7e-5}$ and $\lambda_{\mathrm{Edd}}^{4639} \sim \num{4e-4}$, respectively, after taking into account a bolometric correction from \citet{duras2020a}.
The NSC masses are $\log_{10} \, M_{\star}^{\mathrm{nsc}, \, 2903} \,/\, \mathrm{M}_{\odot} \sim 7.71$ \citep{pechetti2020a} and $\log_{10} \, M_{\star}^{\mathrm{nsc}, \, 4639} \,/\, \mathrm{M}_{\odot} \sim 7.05$ \citep{georgiev2016a}, resulting in mass fractions of $\log_{10} (M_{\mathrm{bh}} \,/\, M_{\star}^{\mathrm{nsc}}) \sim -1.23$ and $-0.4$, respectively.
These values compare well to other literature values, as we show in \Cref{fig:mass_ratio_galmass}.

\begin{figure}
    \centering
    \includegraphics[width=\columnwidth]{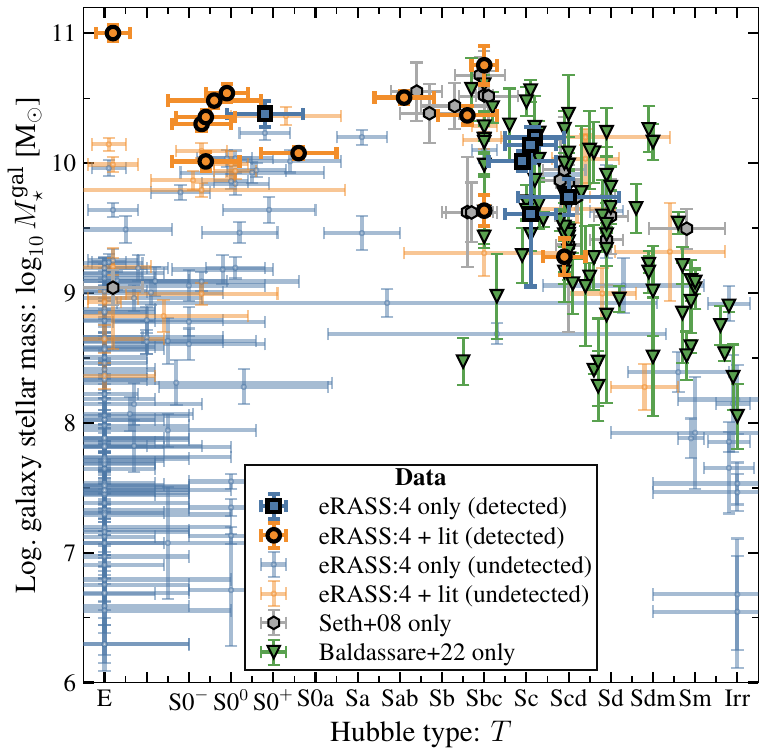}
    \caption{%
        Galaxy stellar mass ($\log_{10} \, M_{\star}^{\mathrm{gal}}$) versus Hubble morphological type ($T$)
        of the X-ray detected (solid blue) and non-detected (fainter blue points) {\erass} data.
        Galaxies, which are part of {\erass} but were also detected by previous work \citep{seth2008a,she2017a,baldassare2022a,ohlson2023a} are shown in orange colour.
        In addition, we show the data by \citet{seth2008a} and \citet{baldassare2022a} with grey hexagons and green triangles, respectively, for galaxies, which are not part of the {\erass} sample.
        The single X-ray detected early-type galaxies at a galaxy stellar mass of approximately \SI{e9}{\Msun} is NGC{\,}4467, whose X-ray properties were analysed by \citet{seth2008a} and \citet{graham2019a}.
    }
    \label{fig:galmass_hubble}
\end{figure}

In comparison to previous studies, our investigation also takes into account a large sample of \num{111} early-type galaxies of various stellar masses.
We show in \Cref{fig:galmass_hubble}, the galaxy stellar mass versus Hubble morphological type plane, with additions of \citet{seth2008a} and \citet{baldassare2022a} for NSCs without observational data in {\erass}.
As previously explored, only the most massive galaxies ($M_{\star}^{\mathrm{gal}} \gtrsim \SI{e10}{\Msun}$) have significant X-ray emission, irrespective of the host galaxy's morphology.
The focus of \citet{baldassare2022a} on the late-type sample of \citet{georgiev2014a} results in detections down to galaxy masses of \SI{\sim e8}{\Msun}, as explored previously.
In contrast, with the exception of one NSC (NGC{\,}4467) from \citet{seth2008a}, no early-type galaxy in the same mass range is X-ray detected within {\erass}.
This could imply several points:
\begin{enumerate}
    \item The accreting MBHs fall below the sensitivity of the cumulative data of {\erosita}.
    \item The black hole occupation fraction is different for galaxies of different Hubble type resulting in fewer X-ray detections at the same host galaxy stellar mass.
    \item Assuming that the massive black hole occupation fraction does not depend on environment, this could indicate smaller black hole masses in these elliptical galaxies, assuming the same Eddington fraction.
    \item While assuming the same massive black hole occupation fraction and typical black hole masses, our results could indicate that the Eddington fraction is different between different morphologies, likely caused by a lack of gas available for accretion in the centres of early-type galaxies.
\end{enumerate}
Regarding the first and last items, we can estimate an upper limit to the Eddington fraction of these objects.
Assuming that we can ignore the contributions of HMXBs, the expected X-ray luminosity from binaries for a galaxy with stellar mass $M_{\star}^{\mathrm{gal}} \sim 10^{8.5} \, \si{\Msun}$ is $L_{X, \, \mathrm{bin.}} \sim \SI{e37}{\erg\per\second}$, which is about a factor ten below the upper limits of {\erass} (see \Cref{fig:lumobs_galmass}).
Using a bolometric correction factor of about ten \citep{duras2020a}, the upper limit on the luminosity of a massive black hole would be $L_{\mathrm{bol}, \, \mathrm{max}} \sim \SI{e39}{\erg\per\second}$.
In galaxies of this mass, we would expect to find MBHs with $M_{\mathrm{BH}} \sim \SI{e5}{\Msun}$ from observational data in early-type galaxies \citep{erwin2012a,reines2015a,capuzzo-dolcetta2017a,greene2020a} and $M_{\mathrm{BH}} \sim 10^{6.5} \si{\Msun}$ from simulations \citep[e.g.][]{spinoso2023a}.
For these MBH masses (\SI{e5}{\Msun}, $10^{6.5} \, \si{\Msun}$), a Bolometric luminosity as quoted above ($L_{\mathrm{bol} ,\, \mathrm{max}} \sim \SI{e39}{\erg\per\second}$) would imply an Eddington fraction of at most \SI{0.01}{\percent}.
This value roughly matches and is one dex higher than he values determined for NGC{\,}2903 and NGC{\,}4639 above, yielding an explanation for why we most likely do not detect these low-luminosity AGN in X-rays, if present.
This also sets an upper limit to the hot gas accretion of such systems.

Regarding the other items, current observational data indicate that MBHs in early-type galaxies are more massive than their counterparts in late-types \citep[see the compilation of][]{greene2020a} but the scaling relations are solely based on measurements in massive galaxies and were extrapolated to the dwarf galaxy regime.
The occupation fraction of MBHs appears to be similar, according to recent X-ray and dynamical results \citep[see, again, the compilation of][]{greene2020a}, making the first and last items of the above list most likely.

\subsubsection{MBH occupation fraction from X-rays}
\label{subsubsec:mbh_occupation_fraction_from_x-rays}

Assuming that all significantly X-ray detected NSCs host an AGN, we can infer the combined occupation fraction of NSCs and AGN.
To gain statistical significance, we add to the {\erass} data the sample of \citet{baldassare2022a}.
For their sample we assume that all galaxies classified as having ``diffuse'' emission are non-detections.

\begin{figure}
    \centering
    \includegraphics[width=\columnwidth]{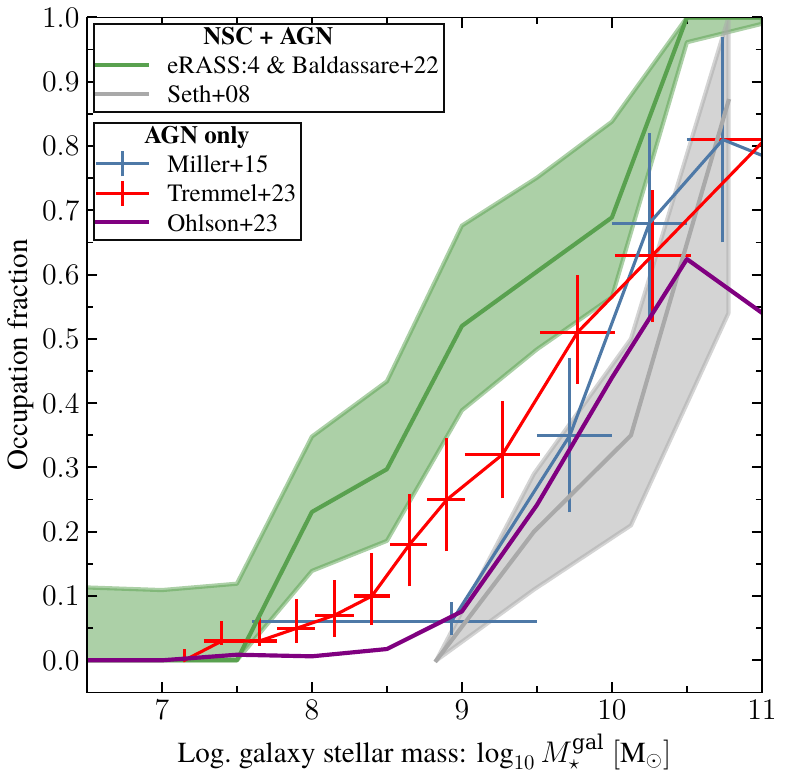}
    \caption{%
        Occupation fraction of nuclear star clusters and active galactic nuclei versus host galaxy stellar mass ($\log_{10} \, M_{\star}^{\mathrm{gal}}$).
        We combine the {\erass} data with the data from \citet{baldassare2022a} to gain statistical significance (green line).
        Literature data for NSCs \& AGN come from \citet[][gray line]{seth2008a}.
        In addition, we show the occupation fractions of AGN from the observational studies of \citet{miller2015a} and \citet[][blue and purple lines, respectively]{ohlson2023a}, and the computational results of \citet[][red line]{tremmel2023a}.
    }
    \label{fig:occ_frac}
\end{figure}

We show the fraction of detected over the total sample as a function of galaxy and NSC stellar mass in \Cref{fig:occ_frac}.
For comparison, we also add the sample occupation fraction of \citet{seth2008a} for NSCs and the AGN occupation fractions (without information of whether an NSC is present) from \citet{miller2015a} and \citet{ohlson2023a} from observations and \citet{tremmel2023a} from a simulation.
We find that above $M_{\star}^{\mathrm{gal}} \sim \SI{e7}{\Msun}$ the combined AGN \& NSC fraction increases and reaches \SI{100}{\percent} around \SI{e10}{\Msun}.
Our data are slightly elevated compared to the data of \citet{seth2008a} and \citet{ohlson2023a}, which could be related different selection effects (\citealp{seth2008a} use optical spectroscopy, radio, and X-ray data to find evidences of the presence of an AGN) of the samples or instrument-related response functions.
Additionally, because of the big half-energy width of {\erosita}'s PSF \citep[half-energy width of about \SI{30}{\arcsec};][]{predehl2021a}, off-nuclear star forming region can contaminate the measurements and result in a too-high estimate of the NSC and AGN fraction.
Given the uncertainties of the data it remains unclear whether the presence of an NSC can enhance the occupation fraction of active galactic nuclei, not taking into account an enhanced rate of tidal disruption events \citep{pfister2020b}.

\subsection{Caveats}
\label{subsec:caveats}

There exist several caveats in the analysis both related to the measured and expected X-ray luminosities.
The spatial resolution of {\erosita} results in an uncertainty on both the central position of the emission (roughly \SI{4}{\arcsecond}) and allows for contamination by off-nuclear sources, thus not guaranteeing that the emission stems from the NSC.
Instead, HMXBs or ULXs may mimic the emission of an accreting massive black hole in NSCs.
Such contaminating sources may still be present in our sample, despite matching it with the ULX sample of \citet{walton2022a} (\Cref{subsec:erosita_observations}), thus requiring follow-up observations with higher spatial resolution facilities like Chandra.

In low-mass galaxies, the above argument is not problematic because the circular aperture typically contains the galaxy's stellar body out to at least one effective radius (\textit{cf}.\ middle and right columns in \Cref{fig:erass_example}).
However, there are several other challenges in this mass range.
As noted by \citet{lehmer2019a}, the scaling relations to track the contribution by HMXBs and LMXBs contain uncertainty in the low-mass regime because of a poorly sampled X-ray luminosity function.
This makes it unclear how to interpret measured X-ray emission in dwarf galaxies in future {\erosita} data releases in case dwarf galaxies (or stacks of dwarf galaxies) become X-ray detected.
Additionally, the scaling relations of \citet{lehmer2019a} apply for the expected X-ray luminosity of the whole galaxy.
However, in most cases, the aperture used to extract the X-ray photometry only covers part of the galaxy (see \Cref{subsec:erosita_observations}), thus overestimating the contamination from binaries.

Additionally, the influence of globular clusters to the X-ray binary contamination remains unclear.
It is well-known that globular clusters efficiently produce LMXBs \citep[e.g.][]{clark1975b,sivakoff2007a,cheng2018a} and that they can heavily influence the X-ray properties of elliptical galaxies \citep[e.g.][]{irwin2005a,lehmer2014a,lehmer2020a} and, to some degree, late-types as well \citep{pfahl2003a,peacock2009a,hunt2023b}.
This effect is especially important in the dwarf galaxy regime where the importance of globular clusters towards the total mass budget of the galaxy increases, as probed by the specific globular cluster frequency\footnote{The specific globular cluster frequency is often calculated as the total number of globular clusters divided by the galaxy's stellar mass, $S_{N} = N_{\mathrm{GC}} \,/\, M_{\star}^{\mathrm{gal}}$.} \citep[e.g.][]{miller2007a,liu2019a,carlsten2022a}.
There also exists some scatter in the specific frequency of dwarf galaxies (see e.g.\ the environmental dependence discussed in \citealt{carlsten2022a}) requiring a detailed investigation of the X-ray luminosity from globular clusters in dwarf galaxies.

Furthermore, what is not taken into account here is the X-ray contribution from binaries within the NSC itself.
This contribution may be similar to globular clusters, especially in dwarf galaxies where the properties of both systems become similar \citep[e.g.][]{fahrion2022b,hoyer2023a}, but it remains somewhat unclear in higher mass NSCs.
Several works have found that denser globular clusters have a higher probability of hosting X-ray binaries \citep[e.g.][]{kundu2002b,jordan2007c,sivakoff2007a,riccio2022a,hunt2023b} and this probability should increase further for NSCs, which are the densest stellar systems known \citep{neumayer2020a}, especially in massive galaxies \citep{pechetti2020a}.
The expected LMXB contribution to the X-ray budget of NSCs is currently unknown and distinguishing them from low-luminosity AGN requires future work.

In summary, there exist several caveats related to both the measured X-ray luminosity and the expected value from X-ray binaries.
Further studies disentangling the contributions from binaries are required for the targets with significant detections, which are high-mass galaxies at present.
The sample size may increase and extend towards the dwarf galaxy regime if future {\erosita} or Athena \citep{nandra2013a} data are added.

\section{Conclusions}
\label{sec:conclusions}

We combined a compilation of galaxies containing a nuclear star cluster (NSC) with {\erosita} {\erass} data to probe X-ray signatures of an accreting massive black hole (MBH) within them.
Using a sample of more than \num{200} nucleated galaxies with overlapping {\erass} data within the footprint of the German {\erosita} Consortium, we find \num{18} significant detections of which one is related to the presence of an off-nuclear ultra-luminous X-ray source.
However, compared to the expected X-ray contamination from both low- and high-mass X-ray binaries, only three galaxies (NGC{\,}2903, 4212, and 4639) have measured luminosities indicative of the presence of an MBH.
Another six galaxies (NGC{\,}2903, 3384, 4321, 4365, 4639, and 4701) have significantly different X-ray luminosities compared to previous archival measurements, which we interpret as indicative of a variable X-ray AGN.
For NGC{\,}4701, we find variability within the {\erass} data set, which could be related to an intrinsic variability or changes in obscuration.
To confirm the nature of these objects, follow-up observations are necessary.

The MBH to NSC stellar mass fraction versus host galaxy stellar mass of the newly identified AGN compares well to other known systems.
By adding X-ray-based black hole mass estimates, we can significantly expand this parameter space towards lower galaxy stellar masses, apparently confirming a drop in the mass ratio around galaxy stellar masses of \SI{e10}{\Msun}.

Assuming that all X-ray detected NSCs above the expected luminosity from X-ray binaries host AGN, we construct an NSC + AGN occupation fraction by adding data from \citet{baldassare2022a} to gain statistical significance.
The resulting curve has higher occupation fraction than the one of \citet{seth2008a} and the AGN only fractions of \citet{miller2015a,tremmel2023a} and \citet{ohlson2023a}.
The differences may be related to instrument-related response functions and the different half-energy widths of the instruments used.

Large-scale surveys, as those carried out by {\erosita} or by Athena in the future, offer a unique view on X-ray emission in dwarf galaxies, covering low-mass early-type galaxies as well, whose X-ray properties had not been investigated previously, with respect to the presence of NSCs.

\begin{acknowledgements}
The authors thank the editor and anonymous referee for their constructive feedback.
N.H.\ is a fellow of the International Max Planck Research School for Astronomy and Cosmic Physics at the University of Heidelberg (IMPRS-HD).
N.H.\ received financial support from the European Union's HORIZON-MSCA-2021-SE-01 Research and Innovation programme under the Marie Sklodowska-Curie grant agreement number 101086388 - Project acronym: LACEGAL.
R.A.\ received support for this work by NASA through the NASA Einstein Fellowship grant No HF2-51499 awarded by the Space Telescope Science Institute, which is operated by the Association of Universities for Research in Astronomy, Inc., for NASA, under contract NAS5-26555.

This work is based on data from eROSITA, the soft X-ray instrument aboard \emph{SRG}, a joint Russian-German science mission supported by the Russian Space Agency (Roskosmos), in the interests of the Russian Academy of Sciences represented by its Space Research Institute (IKI), and the Deutsches Zentrum für Luft- und Raumfahrt (DLR). The \emph{SRG} spacecraft was built by Lavochkin Association (NPOL) and its subcontractors, and is operated by NPOL with support from the Max Planck Institute for Extraterrestrial Physics (MPE). The development and construction of the eROSITA X-ray instrument was led by MPE, with contributions from the Dr. Karl Remeis Observatory Bamberg \& ECAP (FAU Erlangen-Nuernberg), the University of Hamburg Observatory, the Leibniz Institute for Astrophysics Potsdam (AIP), and the Institute for Astronomy and Astrophysics of the University of Tuebingen, with the support of DLR and the Max Planck Society. The Argelander Institute for Astronomy of the University of Bonn and the Ludwig Maximilians Universitaet Munich also participated in the science preparation for eROSITA. The eROSITA data shown here were processed using the eSASS software system developed by the German eROSITA consortium.

This work is based in part on observations made with the Spitzer Space Telescope, which is operated by the Jet Propulsion Laboratory, California Institute of Technology under a contract with NASA.

This work acknowledges funding from the European Research Council (ERC) under the European Union’s Horizon
2020 research and innovation programme (grant agreement No.~865637).

This work used the following software: \texttt{Astropy} \citep{astropy2013a,astropy2018a}, \texttt{dustmaps} \citep{green2018a}, \texttt{Matplotlib} \citep{hunter2007a}, \texttt{NumPy} \citep{harris2020a}, \texttt{Pandas} \citep{mckinney2010a}, \texttt{eSASS} \citep{brunner2022a}, and \texttt{BXA} \citep{buchner2014a}, \texttt{PyXspec} \citep[][,and Craig Gordon and Keith Arnaud]{arnaud1996a}.

The data underlying this article are available in the article and in its online supplementary material.
\end{acknowledgements}

%
%

\bibliographystyle{aa}
\bibliography{./references.bib}
\end{document}